\documentclass[aps,prl,twocolumn,showpacs,preprintnumbers,amsmath,amssymb]{revtex4-1}

 \def\ep{{\epsilon}}

 \def\frac#1#2{{#1\over #2}}

 \def\s{\sqrt}

\def\be{\begin{equation}}
\def\ee{\end{equation}}
\def\ba{\begin{eqnarray}}
\def\ea{\end{eqnarray}}

 \def\f {\frac}
 
 \def\ap{\alpha}

 \def\ddd{\cdot\cdot\cdot}
 \def\no{\nonumber \\}

 \def\ep{\epsilon}

\usepackage{color}
\usepackage{graphicx}
\usepackage{dcolumn}
\usepackage{bm}

\begin{document}

\title{Thermodynamical Property of Entanglement Entropy for Excited States}
IPMU12-0220; YITP-12-99
\author{Jyotirmoy Bhattacharya$^a$, Masahiro Nozaki$^b$, Tadashi Takayanagi$^{b,a}$ and Tomonori Ugajin$^{a,b}$}

\affiliation{$^{a}$Kavli Institute for the Physics and Mathematics of the Universe,\\
University of Tokyo, Kashiwa, Chiba 277-8582, Japan}

\affiliation{$^b$Yukawa Institute for Theoretical Physics,
Kyoto University, \\
Kitashirakawa Oiwakecho, Sakyo-ku, Kyoto 606-8502, Japan}

\date{\today}

\begin{abstract}
We argue that the entanglement entropy for a very small
subsystem obeys a property which is analogous to the first law of
thermodynamics when we excite the system. In relativistic setups,
its effective temperature is proportional to the inverse
of the subsystem size. This provides a universal relationship between the energy and the amount of quantum information. We derive the results using holography and confirm them in two dimensional field theories. We will also comment on an example with negative specific heat and suggest a connection between the second law of thermodynamics and the strong subadditivity of entanglement entropy.
\end{abstract}

\maketitle

{\bf{1. Introduction}} In thermodynamics, when the total energy $E$
of a system is increased, its entropy $S$ grows accordingly. Its
gradient is the definition of temperature $T$ and this leads to the
first law of thermodynamics $dE=TdS$. Since the entropy counts the
number of microstates, this is a fundamental law which relates the
amount of information included in a system to its total energy.

Thus one may wonder if there is an analogous relation for general
quantum systems which are far from equilibrium. One such example
will be a system at zero temperature, i.e. pure state. We can excite
the system, for example, by producing massive particles. It is
well-known that a good measure of quantum information for a pure
state is the entanglement entropy. Therefore, in this letter we will
study how the entanglement entropy for a certain region grows when
we increase its energy. We will largely employ the AdS/CFT
correspondence \cite{Maldacena} and calculate the entanglement
entropy
 holographically \cite{RT}.

{\bf{2. Holographic Entanglement Entropy}} Consider an excited state
in a $d$ dimensional conformal field theory (CFT). We assume it is
almost static and translational invariant. The AdS/CFT
correspondence argues that its ground state is equivalent to gravity
on a $d+1$ dimensional anti de-Sitter space $AdS_{d+1}$
\cite{Maldacena}. The latter is called the gravity dual of the
former. Thus we start with the asymptotically $AdS_{d+1}$
background: \be
ds^2=\f{R^2}{z^2}\left[-f(z)dt^2+g(z)dz^2+\sum^{d-1}_{i=1}(dx_i)^2\right].\label{pads}
\ee Near the boundary $z\to 0$, we can assume $g(z)\simeq
1/f(z)\simeq 1+mz^d,$ where $m$ are constants. We calculate the
energy density $T_{tt}$ of the excited state in the CFT from
(\ref{pads}) by using the holographic energy stress tensor
\cite{Ren}: \be T_{tt}=\f{(d-1)R^{d-1}m}{16\pi G_N}. \label{EMt} \ee
We do not need to make any assumptions about the infrared region
$z\to \infty$ for the argument below. For example, we can have
objects such as black branes or  stars in the infrared region. The
former has  a horizon and is a thermal state, while the latter does
not have any horizon and is dual to a zero temperature state.

To define the entanglement entropy $S_A$ for a subsystem $A$,
we divide the total system into $A$ and $B$ and consider the
reduced density matrix on $A$, called $\rho_A$. $\rho_A$ is
defined by tracing out with respect to $B$: $\rho_A=\mbox{Tr}_B\rho_{tot}$,
where $\rho_{tot}$ is the density matrix for the total system.
The entanglement entropy is defined by $S_A=-\mbox{Tr}\rho_A\log \rho_A$.
In the gravity dual, we can calculate the holographic entanglement entropy by
\be
S_A=\f{\mbox{Area}(\gamma_A)}{4G_N},  \label{HEE}
\ee
where
$\gamma_A$ is a minimal area surface which ends at $z=0$ on the boundary of $A$ \cite{RT}.

Our first choice of subsystem $A$ is a strip defined by $0<x_1<l,\ \
\ -L/2<x_{2,3,\ddd,d-1}<L/2$, where $L$ is taken to be infinite. We
can parameterize the minimal surface $\gamma_A$ by $x_1=x(z)$. Then
its area is computed as \be
\mbox{Area}=2R^{d-1}L^{d-2}\int^{z_*}_{\ep}\f{dz}{z^{d-1}}\s{g(z)+x'(z)^2}.
\ee By minimizing this area functional, we can determine the shape
of $x(z)$. Finally, this leads to \ba &&
\mbox{Area}=2R^{d-1}L^{d-2}\int^{z_*}_{\ep}\f{dz}{z^{d-1}}
\s{\f{g(z)}{1-\left(\f{z}{z_*}\right)^{2(d-1)}}}, \label{areas} \\
&& l=2\int^{z_*}_0 dz \f{z^{d-1}}{z_*^{d-1}}\s{\f{g(z)}{1-\f{z^{2(d-1)}}{z_*^{2(d-1)}}}}\ \ \label{ls},
\ea
where $z=z_*$ is the turning point of $\gamma_A$, i.e. the maximal value of $z$ on $\gamma_A$.

Now, we impose an important assumption in this paper that $l$ is
very small such that \be ml^d<<1. \label{limitr} \ee This means that
$\gamma_A$ is localized near the asymptotically AdS region and this
is the reason why we can ignore the detailed of infrared region. In
this limit, we can expand (\ref{areas}) and (\ref{ls}) up to the
first order perturbation by $ml^d$ and eventually obtain \ba
&& S_A=S^{(0)}_A+\Delta S_A, \label{rest} \\
&& S^{(0)}_A=\f{R^{d-1}L^{d-2}}{2(d-2)G_N}\left[\f{1}{\ep^{d-2}}-\f{2^{d-2}\pi^{\f{d-1}{2}}
\Gamma\left(\f{d}{2(d-1)}\right)^{d-1}}
{\Gamma\left(\f{1}{2(d-1)}\right)^{d-1}\cdot l^{d-2}}\right], \no
&& \Delta S_A=\f{R^{d-1}mL^{d-2}l^2}{32 (d+1)G_N}\cdot \f{\Gamma\left(\f{1}{2(d-1)}\right)^2\Gamma\left(\f{1}{d-1}\right)}{
\s{\pi}\cdot\Gamma\left(\f{1}{2}+\f{1}{d-1}\right)\Gamma\left(\f{d}{2(d-1)}\right)^2}.\nonumber
\ea
$S^{(0)}_A$ is the holographic entanglement entropy in the pure $AdS_{d+1}$ \cite{RT}. Thus $\Delta S_A$ measures how much $S_A$ is increased in the exited state compared with the ground state of the CFT.

On the other hand, the increased amount of energy in the subsystem $A$ is given by
\be
\Delta E_A=\int dx^{d-1}\Delta T_{tt}=\f{(d-1)mlL^{d-2}R^{d-1}}{16\pi G_N}.
\ee

Therefore we find the following relation
\be
\f{\Delta S_A}{\Delta E_A} =\f{\s{\pi}\Gamma\left(\f{1}{2(d-1)}\right)^2\Gamma\left(\f{1}{d-1}\right)}
{2(d^2-1)\Gamma\left(\f{1}{2}+\f{1}{d-1}\right)\Gamma\left(\f{d}{2(d-1)}\right)^2}\cdot l
\label{stripeeth}
\ee

For another example, consider the case where $A$ is given by a round
ball with radius $l$: $\sum_{i=1}^{d-1}x_i^2\leq l^2$. Its minimal
surface $\gamma_A$ is specified by $r=r(z)$. The area is computed as
\ba \mbox{Area}&=&R^{d-1}\Omega_{d-2}\int^u_\ep \f{dz}{z^{d-1}}\cdot
r(z)^{d-2}\s{g(z)+r'(z)^2},\nonumber \ea where $\Omega_{d-2}=2\cdot
\f{\pi^{\f{d-1}{2}}}{\Gamma\left(\f{d-1}{2}\right)}$ is the volume
of $S^{d-2}$ with the unit radius.

By solving the equation of motion for $r(z)$ derived from the above
area functional, we find the following solution up to the first
order of $ml^d$ \be
r(z)=\s{u^2-z^2}+m\f{2u^{d+2}-z^d(u^2+z^2)}{2(d+1)\s{u^2-z^ 2}}, \ee
where we assumed a regularity at $z=u$. The parameter $u$ is a free
positive constant and is related to the radius $l$ of the subsystem
$A$ by $l=r(0)=u+m u^{d+1}/(d+1).$

Then, the area is rewritten as \ba &&
\mbox{Area}=R^{d-1}\Omega_{d-2}\int^u_\ep dz \f{u}{z^{d-1}}
(u^2-z^2)^{\f{d-3}{2}}\cdot \left(1+mK(z)\right),\no &&
K(z)\equiv\f{2(d-2)u^{d+2}\!-\!2u^dz^2\!+\!(3+d)u^2z^d\!-\!3(d-1)z^{d+2}}
{2(d+1)(u^2-z^2)}. \nonumber \ea

In the end, we find \be \Delta
S_A=\f{\pi^{\f{d-1}{2}}}{4(d+1)\Gamma\left(\f{d-1}{2}\right)} \cdot
\f{R^{d-1}}{G_N}\cdot ml^d.  \label{deltas} \ee

By computing $\Delta E_A$ from (\ref{EMt}), we finally obtain \be
\f{\Delta S_A}{\Delta E_A}=\f{2\pi}{d+1}\cdot l. \label{ther} \ee

If we take the limit $(\ref{limitr})$, the entanglement entropy in
CFTs, as shown in (\ref{stripeeth}) and (\ref{ther}), satisfies a
universal relation analogous to the first law of thermodynamics \be
T_{ent}\cdot \Delta S_A=\Delta E_A, \label{firstlaw} \ee where the
effective temperature (`entanglement temperature') $T_{ent}$ is
proportional to the inverse of $l$: \be T_{ent}=c\cdot l^{-1}.
\label{etem} \ee $c$ is an order one constant. When the subsystem
$A$ is a round sphere, we find $c=\f{d+1}{2\pi}$.

In the field theoretic language, this argues that in a strongly
coupled large $N$ gauge theories, the relation (\ref{firstlaw}) and
(\ref{etem}) are satisfied if we take the subsystem size $l$ to be
very small such that \be T_{tt}\cdot l^d<< R^{d-1}/G_N\sim
O(N^2). \label{condf} \ee

\begin{figure}[t]
   \begin{center}
       \includegraphics[height=3cm]{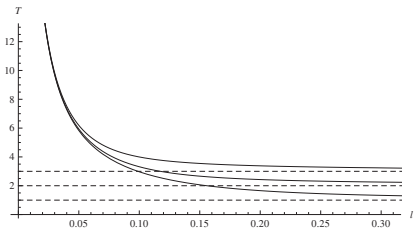}
      \end{center}
   \caption{The effective temperature
   $T_{ent}=\f{d(\Delta E_A)}{d(\Delta S_A)}|_{l=fixed}$ as
   a function of $l$ based on the holographic
   calculation using the AdS$_4$ black hole at temperature $T$.
   We set $T=1,2,3$.}\label{fig:ltplot}
\end{figure}

What we learn from (\ref{firstlaw}) is the universal statement that the
 amount of information included in a small subsystem $A$ is proportional to
  the energy included in $A$. The AdS/CFT predicts that the constant $c$ in
   (\ref{etem}) is universal when we fix the shape of the subsystem $A$.
   Note that the source of the excitation energy is arbitrary.
   It can be a temperature increase or can be creations of massive objects at
   zero temperature. The condition (\ref{condf}) is crucial here. If we do not assume
   this, then $T_{ent}$ defined by (\ref{firstlaw}) is no longer universal and
   depends on the details of the excitations. In the gravity side, the
   result depends on the details of the infrared region. For
   example, in the AdS Black hole at temperature $T$,
   $T_{ent}$ defined by (\ref{firstlaw}) approaches
(\ref{etem}) in the limit $l\to 0$, while $T_{ent}$ becomes $T$ in
the opposite limit $l\to \infty$ as plotted in Fig.\ref{fig:ltplot}.

   Finally we would like to mention that there are earlier works \cite{MSK,CHM}, where the entanglement entropy for ground states $S^{(0)}_A$ was interpreted as thermal entropy.

{\bf{3. Comparison with QFT at Finite $T$}} Since we analyzed the
gravity duals so far, it is useful to turn to a direct calculation
in quantum field theories (QFTs). However, at present, this has only
been done for two dimensional CFTs. First we know the general
formula of $S_A$ \cite{CaCa} at finite temperature $T=\beta^{-1}$
when $A$ is an interval of width $l$: \be S_A=\f{c}{3}\log\left(
\f{\beta}{\pi\ep}\sinh\f{\pi l}{\beta} \right). \label{ftee} \ee By
expanding this in the limit $l<<\beta$, we find \be \Delta
S_A=\f{c\pi^2 T^2l^2}{18}=\f{R ml^2}{48G_N}=\f{\pi}{3}l\cdot\Delta
E_A, \ee where we employed the standard relation $c=\f{3R}{2G_N}$
\cite{BrHe} and $m=(2\pi T)^2$ in $AdS_3/CFT_2$. This agrees with
our holographic calculation in (\ref{rest}). Moreover, if we set the
width of interval is $2l$, then it also agrees with (\ref{ther}).

Also $\Delta S_A$ was recently calculated for low-energy excitations
on a cylinder with the circumference $L_{cy}$ \cite{UAM} \be \Delta
S_A=\f{2\pi^2}{3}(h+\bar{h}) \f{l^2}{L^2_{cy}}=\f{\pi}{3}l\cdot
\Delta E_A, \ee where $(h,\bar{h})$ are the chiral and anti-chiral
conformal weights of a primary operator.

In higher dimensional CFTs, we were not aware of relevant QFT
calculations at present. Instead, our result (\ref{stripeeth})
presents predictions: $\Delta S_A\propto l^2 L^{d-2}T^d$ at
temperature $T$ with $lT<<1$; $\Delta S_A\propto \Delta E_A\cdot
l^d$ for a CFT$_d$ on $R\times S^d$ (unit radius) with the subsystem
radius $l<<1$, where $\Delta E_A$ is the conformal weight of the
excites state.

{\bf{4. Inhomogeneous Backgrounds}}
We can generalize our calculation of $\Delta S_A$
to inhomogeneous backgrounds. As an example, consider
the gravity dual:
\ba
&& ds^2=\frac{R^2}{z^2}\left[-f(z)dt^2+g(r,z)dz^2+dr^2+r^2d\Omega_{d-1}\right], \no
&& g(r,z)=1 + m (1 +a r +b r^2) z^d.\nonumber
\ea
This is dual to a $d$ dimensional CFT with $r$ dependent energy density $T_{tt}=\f{(d-1)R^{d-1}m}{16\pi G_N}(1 +a r +b r^2)$. We calculate $\Delta S_A$ when $A$ is a round sphere (radius $l$). The results for $d=2,3,4$ are given by
\ba
&& \Delta S_A=\frac{ l^2 m R }{240G_N}\left\{20+3 l (5 a+4 b l)\right\}\ \ \ (d=2),\no
&&  \Delta S_A=\frac{\pi m l^3 R^2 }{240 G_N}\left[15+2 l (6 a+5 b l)\right]\ \ \ (d=3),\no
&&  \Delta S_A=\frac{\pi m l^4 R^3 }{420 G}\left[42+5 l (7 a+6 b l)\right]\ \ \ (d=4).
\ea
In this way, we can extract the details of the metric from the behavior of
the entanglement entropy.

{\bf{5. Hyperscaling Violating Backgrounds}} We can extend our
analysis to gravity duals of more general non-relativistic critical
points called the hyperscaling violating geometry: \cite{HS}: \be
ds^2\!=\!\f{R^2}{r^2}\!\left(\!-
r^{-\f{2(d-1)(z-1)}{d-1-\theta}}dt^2\!+\!
r^{\f{2\theta}{d-1-\theta}}dr^2 \!+\!dx_i^2\!\right)\!, \label{hsv}
\ee
where $\theta$ is the hyperscaling violating exponent and $z$ is the
dynamical exponent. We can realize this geometry as a solution in
Einstein-Maxwell-scalar theory \cite{HS}.

We heat up this system at temperature $T$ and consider the
holographic entanglement entropy for the strip subsystem with the
width $l$. Strictly speaking, we need to embed everything into an
asymptotic AdS space, where the IR geometry looks like (\ref{hsv}).
We define $z_1$ to be the scale where the hyperscaling violating
geometry starts to appear. We are interested in the region
$z_1<<l<<T^{-1}$. In this case, the
analysis of the holographic entanglement entropy is not affected by
the presence of the asymptotically AdS$_4$ region and we can
focus on the metric (\ref{hsv}).

The finite temperature solution is the black brane solution given by
multiplying $f(r)$ and $1/f(r)$ in front of $dt^2$ and $dr^2$ in (\ref{hsv}),
where $f(r)=1-(r/r_H)^{(d-1)\left(1+\f{z}{d-1-\theta}\right)}.$ The
horizon is situated at $r=r_H$. It is immediate to find that the
thermal entropy as the horizon
 area and then by using the first law of thermodynamics we can
 calculate the energy density. We can calculate $\Delta S_A$
as in the previous analysis (\ref{rest}). The
final results look like $\Delta S_A/\Delta E_A=l^z/c_{h},$ where
$c_{h}$ is an order one constant. Thus we find $T_{ent}\propto
l^{-z}$, which is natural from the definition of the dynamical
exponent $z$.

{\bf{6. Negative/Positive Specific Heat}}
We would like to discuss an entanglement entropy counterpart
of the positivity of specific heat, which is required by the second law
of thermodynamics. A good example of this kind is the D3-brane shell
\cite{KLT}. The near horizon geometry of D3-brane shell is given by
the metric \ba && ds^2=\f{R^2}{z^2h(z)}\left(\sum_{\mu=0}^3dx_\mu
dx^\mu\right)+R^2 h(z)\left(\f{dz^2}{z^2}+d\Omega_5^2\right),\no
&& h(z)=1\ \ (z\leq z_0),\ \  \ h(z)=z_0^2/z^2\ \
(z\geq z_0).
\ea
This solution represents D3-branes distributed
at $z=z_0$ like a spherically symmetric shell. The geometry for
$z>z_0$ is a flat spacetime $R^{1,9}$, while that for $z<z_0$ is the
AdS$_5\times$ S$^5$. This is dual to a $N=4$ super Yang-Mills in the
Coulomb branch \cite{KLT}.

By exciting this system, we can make a small black hole
at $z=\infty$. This is a Schwarzchild black 3-brane in flat
spacetime and thus it has the negative specific heat. This solution
is thermodynamically unstable and will finally decay into the
standard AdS black hole solution.

\begin{figure}[t]
   \begin{center}
       \includegraphics[height=3cm]{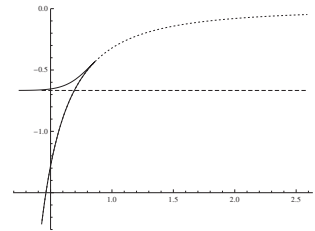}
      \end{center}
   \caption{  The plot describes the regularize minimal surface area
   (divided by $\pi^3R^8L^2$) as a function of the width $l$ in the D3-brane shell.
    We choose $z_0=1$. The two thick curves correspond to the connected minimal surfaces,
    where upper one is not concave and thus is not physical. The horizontal dashed line represents the disconnected surface in the D3-brane shell.
    The dotted curve is the result for the pure
    AdS.}\label{fig:negative}
\end{figure}

Consider the holographic entanglement entropy for the strip with the
width $l$ in the D3-brane shell. It is computed by finding the 8
dimensional minimal area surface $x=x(z)$ at a time $t=0$. The area
functional is given by \be \mbox{Area}=2\pi^3R^8L^2
\int^{z_*}_{0}\f{dz}{z^3}h(z)\s{x'(z)^2+h(z)^2}, \ee where $z=z_*$
is the turning point of a connected surface. There
is another candidate of the minimal surface which consists of two
disconnected surfaces defined by $x=l/2$ and $x=-l/2$. We plotted
the regularized area, which is defined by subtracting the area law
divergence $\f{L^2}{\ep^2}$ \cite{area} from the area,
in Fig.\ref{fig:negative}. The result
 looks similar to the one in the gravity duals of confining
gauge theories \cite{Soliton}. Notice that there are two branches in
the connected surface for a fixed value of $l$. The one with lower
area is sensible as it satisfies the strong subadditivity equivalent
to the concavity $\f{d^2 S_A}{d l^2}\leq 0$ \cite{LR,CHS,HT}.
However, the other one is not. Since we need the smallest area when
there are several candidates of $\gamma_A$, we choose the
lowest area surface. For a certain value of $l=l_c$, the disconnected
surface is favored as it has a smaller area, where
$S_A$ shows an analogue of phase transition as a function of $l$.

Next let us turn to the unstable surface which is not concave.
Even though this surface does not contribute to $S_A$,
 it is intriguing to understand the meaning of its presence.
This surface extends deep into the IR flat space region $z\to\infty$.
Thus, it is natural to expect that this is related to a Schwartzshild black 3-brane in
$R^{1,9}$. This makes us to suspect that the negative specific
heat is related to the violation of strong subadditivity.
Remember that the entropy $S$ and
energy $E$ of a black $p$-brane in $R^{1,9}$ behave like
 $S_{p}\propto V_{p}T^{p-8}$ and $E_p\propto
V_{p}T^{p-7}$, where $V_p$ is the volume of the $p$-brane and $T$ is
its temperature. On the other hand, in the small $l$ limit, we can
show the finite part of holographic entanglement entropy behaves as
$[S_A]_{finite}\propto\f{R^8L^2l^6}{z_0^8}$. By identifying $V_p$
with the volume of subsystem $A$ and the temperature $T$ with $1/l$,
we find that $[S_A]_{finite}$ agrees with the entropy of black
3-brane $S_3$ up to a numerical factor.

Finally, let us study the connection between the sign of specific
heat and the strong subadditivity. We assume that the finite part of
the entanglement entropy in any dimension behaves like
$[S_A]_{finite}=\ap\cdot  l^{q+1}$ for the strip subsystem $A$ with
the width $l$. Since we expect $S_A$ is a monotonically increasing
function of $l$, $\ap$ is positive (or negative) when $q+1>0$ (or
$q+1<0$). The strong subadditivity (i.e. concavity) requires $q\leq
0$. By identifying the temperature and volume as $T\propto 1/l^z$
and $V_p\propto lL^{p-1}$, we can relate $S_A$ to the thermal
entropy $S_{th}\propto V_p\cdot T^{-q/z}$. The positive specific
heat requires $q\leq 0$. In this way, we find that the strong
subadditivity is equivalent to the positivity of specific heat in
this setup.

{\bf{7. Discussion}} In conclusion, the main result of this paper is
that the entanglement entropy in CFTs satisfies the first law like
relation (\ref{firstlaw}) with the universal `entanglement
temperature' (\ref{etem}) when the subsystem $A$ is so small that
(\ref{condf}) is satisfied. This means that the variation $\Delta
S_A$ is given by physical observables. We derived this from the
AdS/CFT and confirmed this in 2d CFTs. An interesting future problem
is to check this directly in higher dimensional QFTs.

There are many different ways to add the energy $\Delta E_A$ to the
subsystem $A$. Consider a zero temperature setup where $\Delta E_A$ depends on an external
parameter $x$ such as the distance between two interacting particles.
Then we can express its force
$F_x$ in terms of $\Delta S_A$ as follows \be
F_x=-\f{d}{dx}\Delta E_A(x)=-T_{ent}\cdot\f{d}{dx}\Delta S_A(x). \label{ef} \ee Though
this might look like an entropic force \cite{Verlinde,Fur}, the sign is
opposite. This force $F_x$ acts so that it tries to reduce the
entanglement entropy. In the gravity dual, $F_x$ can be
 a gravitational force or some other forces which exist in the gravity theory.
  It is an interesting future problem to study the implication of (\ref{ef}).

{\bf Acknowledgements} We thank D.~Fursaev, C.~Herzog, E.~Lopez,
S.~Matsuura, R.~Myers, T.~Nishioka and S.~Trivedi for useful
discussions. TT is supported by JSPS Grant-in-Aid for Challenging
Exploratory Research No.24654057. TU is supported by the JSPS
fellowship. JB, TT and TU are supported by World Premier
International Research Center Initiative (WPI Initiative) from the
Japan Ministry of Education, Culture, Sports, Science and Technology
(MEXT).

\end{document}